\documentclass[aps,prb,twocolumn]{revtex4}
\usepackage{graphicx}
\usepackage{bm}
\begin{document}
\title{Oscillatory nonlinear differential magnetoresistance of highly mobile 2D electrons\\ 
in high Landau levels}
\author{X. L. Lei}
\affiliation{Department of Physics, Shanghai Jiaotong University,
1954 Huashan Road, Shanghai 200030, China}

\begin{abstract}

We examine the current-induced magnetoresistance oscillations 
in high-mobility two-dimensional electron systems using the balance-equation scheme
for nonlinear magnetotransort.  
The reported analytical expressions for differential magnetoresistivity 
at high filling factors in the overlapping Landau-level regime, which 
show good agreement with the experimental observation and the numerical calculation,
may be helpful in extracting physical information from experiments.

\end{abstract}


\maketitle

In addition to the universally existing Shubnikov-de Haas oscillations (SdHO),
many different kinds of magnetoresistance oscillations were discovered in the past few years
in high-mobility two-dimensional (2D) electron systems (ES) subject to a weak perpendicular 
magnetic field and have become a field of great interest. 
These resistance oscillations always occur at low temperatures and are related to magnetotransport 
of 2D electrons occupying high Landau levels (LLs). Among them
the microwave-induced magnetoresistance oscillations and the related zero-resistance states
were the central focus of most experimental\cite{Zud01,Mani02,Zud03,Dor03,Mani04,Willett,
Dor05,Stud,WZhang07,Hatke08} and theoretical\cite{Ryz03,Durst,Lei03,
Vav04,Dmitriev03,DGHO05,Torres05,Ina-prl05,Kashuba,Lei07-2,Khodas08} studies. 
Recently, the oscillatory behavior in the nonlinear magnetotransport 
has attracted much attention: in a 2D system even without irradiation, a relatively weak 
current can induce drastic suppression and strong oscillations of the differential 
magnetoresistance, and may result in a state of zero-differential resistance.\cite{Yang02,Bykov,JZhang07-1,WZhang07-1,WZhang08-1} 

Several theoretical models have been proposed in an attempt to explain this interesting 
nonlinear phenomenon.\cite{Lei07-1,Vav07,Auerbach07,Kun08}
Numerical examinations based on the current-control transport scheme were shown 
in good agreement with the experimental observation for differential magnetoresistivity
as a function of the ratio of the current density to the magnetic field
in both the magnetic-field sweeping and the current-sweeping configurations
covering both separated and overlapping Landau level regimes.\cite{Lei07-1}  

It is reported recently that by a systematic analysis 
of current-induced magnetoresistance oscillations, important physical
information about electron-electron interaction on the single particle life time
can be extracted.\cite{Hatke09-4} 
From the point of view of experiment, an analytical expression for differential
magnetoresistivity, even applies only within limited ranges, 
is highly desirable because it can be of great help to extract important physical information
from experimental data.
So far, a reliable analytical expression derived 
from experimentally confirmed theoretical models is still lacking.

 We examine this issue 
 based on the balance-equation scheme of nonlinear magnetotransport,\cite{Lei07-1}
which deals with a 2D system consisting of $N_{s}$ electrons in a unit area of 
the $x$-$y$ plane and subjected to a uniform magnetic field ${\bm B}=(0,0,B)$ 
in the $z$ direction.  These electrons, scattered 
by randomly distributed impurities and by phonons in the lattice,
perform an integrative drift motion 
under the influence of a uniform electric field ${\bm E}$ in the $x$-$y$ plane.
For high mobility and high carrier-density systems 
in which effects of electron-impurity and electron-phonon scatterings 
are weak in comparison with their internal thermalization, 
the steady transport state at lattice temperature $T$ is described by 
the electron average drift velocity ${\bm v}$ and
an electron temperature $T_{\rm e}$. They satisfy the following force- and energy-balance
equations:
\begin{eqnarray}
N_{s}e{\bm E}+N_{s} e ({\bm v} \times {\bm B})+
{\bm f}({\bm v})&=&0,\label{eqforce}\\
{\bm v}\cdot {\bm f}({\bm v})+ w({\bm v})&=&0. \label{eqenergy}
\end{eqnarray}
Here ${\bm f}({\bm v})={\bm f}_{\rm i}({\bm v})+{\bm f}_{\rm p}({\bm v})$ 
is the damping force against the electron drift motion due 
to impurity and phonon scatterings respectively,
\begin{eqnarray}
&&{\bm f}_{\,\rm i}=\sum_{{\bm q}_\|}| U({\bm q}_\|)| ^{2}
{\bm q}_\|\,{\Pi}_{2}
\big({\bm q}_\|,\, {\bm q}_\|\cdot\! {\bm v} \big),\label{exfi}\\
&&{\bm f}_{\rm p}=2 \sum_{{\bm q},\lambda}| M({\bm q},\lambda)| ^{2}
{\bm q}_\|\,{\Pi}_{2}
\big({\bm q}_\|,\,\Omega_{{\bm q}\lambda}\!+{\bm q}_\|\!\cdot {\bm v}\big)\nonumber\\
&&\hspace{1.5cm}\times \left[ n \!\left( \frac{\Omega_{{\bm q}\lambda}}{T} \right)- 
n \!\left(\frac{\Omega_{{\bm q}\lambda}\!+{\bm q}_\|\!\cdot {\bm v}}{T_{\rm e}} \right)\right],
\label{exfp}
\end{eqnarray}
and $w({\bm v})$ is the electron energy-loss rate to the lattice due to electron-phonon 
interactions having an expression obtained from the right-hand side of Eq.\,(\ref{exfp}) 
by replacing the ${\bm q}_\|$  factor with 
$\Omega_{{\bm q}\lambda}$, the energy of a wavevector-${\bm q}$ phonon in branch $\lambda$.
In these equations, 
$n(x)=1/({\rm e}^x-1)$ is the Bose function,
$U({\bm q}_\|)$ is the effective impurity potential,
$M({\bm q},\lambda)$ is the effective electron-phonon coupling matrix element,
and ${\Pi}_2({\bm q}_\|,{\omega})$ 
is the imaginary part of the electron density correlation function 
at electron temperature $T_{\rm e}$ in the presence of the magnetic field. 

The density correlation function ${\Pi}_2({\bm q}_{\|}, {\omega})$  
of a 2D electron gas in the magnetic field can be written in the Landau representation as
\begin{eqnarray}
&&\hspace{-0.7cm}{\Pi}_2({\bm q}_{\|},{\omega}) =  \frac 1{2\pi
l_{B}^2}\sum_{n,n'}C_{n,n'}(l_{B}^2q_{\|}^2/2)\, 
{\Pi}_2(n,n',{\omega}),
\label{pi_2q}\\
&&\hspace{-0.7cm}{\Pi}_2(n,n',{\omega})=-\frac2\pi \int d\varepsilon
\left [ f(\varepsilon )- f(\varepsilon +{\omega})\right ]\nonumber\\
&&\,\hspace{2.3cm}\times\,\,{\rm Im}G_n(\varepsilon +{\omega})\,{\rm Im}G_{n'}(\varepsilon ),
\label{pi_2ll}
\end{eqnarray}
where $l_{B}=\sqrt{1/|eB|}$ is the magnetic length,
$
C_{n,n+l}(Y)\equiv n![(n+l)!]^{-1}Y^l{\rm e}^{-Y}[L_n^l(Y)]^2
$
with $L_n^l(Y)$ the associate Laguerre polynomial, $f(\varepsilon
)=\{\exp [(\varepsilon -\varepsilon_{F})/T_{\rm e}]+1\}^{-1}$ is the Fermi 
function at electron temperature $T_{\rm e}$ with $\varepsilon_{F}$
the Fermi level of the system, 
and ${\rm Im}G_n(\varepsilon )$ is the density-of-states (DOS) function of the broadened LL $n$.

The LL broadening depends on impurity, phonon and electron-electron scatterings.
In a high-mobility GaAs-based 2D system the dominant elastic scattering comes
from impurities or defects in the background or close proximity,\cite{Umansky,Yang02} 
and phonon and electron-electron scatterings are 
also not long-ranged because of the screening. 
The correlation lengths of these scattering potentials are much smaller than
the cyclotron radii $R_c$ of electrons involving in the transport subject to a weak magnetic field,
but much larger than $R_c/n$ for very high ($n\gg 1$) LLs.
 In this case, the broadening of the LL is expected  
to be a Gaussian form\cite{Raikh-1993} 
\begin{equation}
{\rm Im}G_n(\varepsilon)=-(2\pi)^{\frac{1}{2}}{\Gamma}^{-1}
\exp[-2(\varepsilon-\varepsilon_n)^2/{\Gamma}^2].
\label{gauss}
\end{equation}
In this, 
$\varepsilon_n=(n+\frac{1}{2})\omega_c$ is the center of the $n$th LL,
$\omega_c=eB/m$ is the cyclotron frequency with $m$ the effective mass of the electron,
and $\Gamma$, the half-width of the LL, is $B^{1/2}$-dependent expressed as\cite{Coleridge-97} 
\begin{equation}
{\Gamma}=(2\omega_c/\pi \tau_s)^{1/2}, 
\label{gamma12}
\end{equation} 
where $\tau_s$ is the single-particle lifetime or quantum scattering time of the electron
in the zero magnetic field. This single-particle life time $\tau_s$, 
relating to impurity, phonon and electron-electron scatterings,
is generally temperature dependent. 

The DOS function (\ref{gauss}) for the $n$th LL is valid and can be used    
in both separated and overlapping LL regimes.
The total DOS (double spins) for a 2D system of unit area
in the presence of a magnetic field is given by
$g(\varepsilon)= -\sum_n {\rm Im}G_n(\varepsilon)/({\pi^2l_{B}^2})$,
and the Fermi level $\varepsilon_{F}$
is determined by the electron sheet density $N_{s}$ from the equation
$\int d\varepsilon f(\varepsilon)g(\varepsilon)=N_{s}$.
In the case of high LL filling  $\varepsilon_{F}$ is essentially the same 
as that of a 2D electron gas having the same carrier density $N_{s}$ without magnetic field, 
$\varepsilon_{F}=k_{F}^2/2m$
($k_{F}$ is the Fermi wavevector),
and $\nu \equiv \varepsilon_{F}/\omega_c$ is the the filling factor.

When the LL width $2{\Gamma}$ is larger than the level spacing $\omega_c$ 
(overlapping LLs), the Dingle factor 
\begin{equation}
\delta=\exp\left(-{\pi }/{\omega_c}\tau_s \right)
\end{equation}
is smaller than 0.3.
Keeping only terms of the lowest order in $\delta$ or of the fundamental harmonic oscillation,
we have the following approximate expression for the DOS:
\begin{equation}
 g(\varepsilon)\approx \frac{m}{\pi}\big[1-2\delta
\cos\left(2\pi \varepsilon/\omega_c\right)\big].
\label{dg-g}
\end{equation}

For an isotropic system where the frictional force ${\bm f}({\bm v})$ is in the opposite direction to 
the drift velocity ${\bm v}$,
 we can write ${\bm f}({\bm v})=f(v){\bm v}/v$.  
In the Hall configuration with the velocity ${\bm v}$ in the $x$ direction
${\bm v}=(v,0,0)$ or the current density $J_x=J=N_{s}ev$ and $J_y=0$,
Eq.\,(\ref{eqforce}) yields, at a given $v$,  
the transverse resistivity $R_{yx}=B/N_{s}e$, and the longitudinal resistivity 
 and differential magnetoresistivity as
\begin{eqnarray}
&&R_{xx}= -f(v)/(N_{s}^2e^2v),\\
&&r_{xx}= -(\partial f(v)/\partial {v})/(N_{s}^2e^2).
\end{eqnarray}

These formulas, expressing the nonlinear resistivity as a function of the drift velocity ${v}$
without invoking electric field, are convenient to direct relate to experiments
where transport measurements are performed by controlling the current.
This is the basic feature of the balance equation approach,\cite{Lei-85} 
in which the drift velocity ${\bm v}$ of the electron system, rather than the electric field, 
plays as the fundamental physical quantity to affect electron transport. In this approach individual (relative)
electrons, treated in the reference frame moving at velocity ${\bm v}$,  
do not directly feel a uniform electric field.
The role of the drift velocity ${\bm v}$ is to provide the electron of wavevector ${\bm q}_{\|}$ with
an energy ${\bm q}_{\|}\cdot{\bm v}$ during its transition from a state to another state induced by impurity or phonon
scattering. This results in an extra frequency ${\bm q}_{\|}\cdot{\bm v}$ in the density correlation function
$\Pi_2({\bm q}_{\|}, \omega+{\bm q}_{\|}\cdot{\bm v})$ in Eqs.(3) and (4). Since in a magnetic field 
the density correlation function is frequency periodic,  
$\Pi_2({\bm q}_{\|},\omega)\sim \Pi_2({\bm q}_{\|},\omega+\omega_c)$, due to periodical LLs,
change of the drift velocity ${\bm v}$ will lead to oscillation of
related physical quantities. At low temperature and high LL filling, 
in view of $\Pi_2({\bm q}_{\|}, \omega)$ function sharply peaking around $q_{\|} \simeq 2 k_{F}$,
the effect of a finite velocity $v$, after the $q_{\|}$ integration, is equivalent to shift the frequency 
in the $\Pi_2$ function an amount of $\omega_j \equiv 2k_{F} v$. Thus when $\omega_j$ varies by a value of $\omega_c$, 
the resistivity experiences change of an oscillatory period.
It is thus convenient using a dimensionless parameter
\begin{equation}
\epsilon_j\equiv\frac{\omega_j}{\omega_c}=\frac{2mk_{F}v}{eB}
=\sqrt{\frac{8\pi}{N_{s}}}\frac{m}{e^2}\frac{J}{B}
\end{equation}
to demonstrate the behavior of nonlinear magnetoresistivity oscillation.

The numerical examination of 
impurity-related differential resistivity\cite{Lei07-1}
shows that 
$r_{xx}^{\rm i}=-(\partial f_{\rm i}(v)/\partial {v})(N_{s}^2e^2)$
oscillates with changing $\epsilon_j$
exhibiting an approximate period $\Delta \epsilon_j \approx 1$,
with maxima at positions somewhat lower than integers $\epsilon_j=n$ ($n=1,2,3,...$)
and minima somewhat lower than half integers $\epsilon_j=n+1/2$ ($n=1,2,3,...$).
The predicted results covering both separated and overlapping Landau level regimes,
are in good agreement with the experimental observation of Ref.\,\onlinecite{WZhang07-1} 
in both magnetic-field sweeping and current-sweeping cases.

Simple and accurate analytical expressions of the impurity-induced resistivity 
$r_{xx}^{\rm i}$
can be derived for short-range potentials in the case of high filling factor $\nu$ 
within the overlapping LL regime.
 
At temperature $T\ll \varepsilon_{F}$ the impurity-induced 
linear ($v\rightarrow 0$)  
resistivity $r_{xx}^{\rm \,i}(0)$ is given, 
to the lowest nonzero order in $\delta$ and the fundamental SdHO part, by
\begin{equation}
r_{xx}^{\rm \,i}(0)=R_{\rm i0}\left[1+2\delta^2 -4 \delta 
D(X) \cos(2\pi \nu)\right],
\label{sdhi}
\end{equation}
where $X\equiv {2\pi^2 T}/{\omega_c}$, $D(X)\equiv X/\sinh(X)$, and $R_{\rm i0}$ 
is the low-temperature ($T \ll \varepsilon_{F}$) linear resistivity of the 2D electron gas
in the absence of magnetic field,\cite{Lei851}
which is directly related to the transport scattering time $\tau_{\rm tr}$ or 
the linear mobility $\mu_0$ as
$R_{\rm i0}=m/(N_{s} e^2\tau_{\rm tr})=1/(N_{s} e \mu_0)$. 
The expression (\ref{sdhi}) shows that $r_{xx}^{\rm \,i}(0)$, though oscillating strongly
with changing filling factor $\nu$, is always positive.
Both its oscillatory (SdHO) and non-oscillatroy parts
 are $\tau_s$-dependent through the $\delta$-involved terms 
in the presence of a magnetic field and may change with temperature. 
The linear resistivity $R_{\rm i0}$ of a 2D electron system without magnetic field, however,
is not affected by its single particle lifetime $\tau_s$ and thus temperature independent.

The impurity-related SdHO [the last term in Eq.\,(\ref{sdhi})] 
exhibits minima at integer filling factor $\nu$ and
quickly diminishes with rising temperature $T\gg \omega_c/2\pi^2$
due to the temperature-dependent factor $D(X)$.
The temperature dependence of the non-SdHO part of the impurity-induced linear resistivity 
of the 2D electron system in the presence of a magnetic field can appear 
mainly through the temperature change of its single particle lifetime $\tau_s$
entering the Dingle-factor coefficient  $2\delta^2$.  

In the case of overlapping LLs and high filling factor $\nu$,
the nonlinear longitudinal differential resistivity $r_{xx}^{\rm i}$ can be expressed, 
at temperature $T_{\rm e}\ll \varepsilon_{F}$, as 
\begin{eqnarray}
&&\hspace{-0.8cm}r_{xx}^{\rm \,i}=R_{\rm i0}\Big\{1+\,2\delta^2\, G(2\pi \epsilon_j)\nonumber\\
&&\hspace{1cm} -4 \delta \,
D(X_{\rm e}) \cos(2\pi \nu) S(2\pi\epsilon_j) \Big\},
\label{nrxxi}
\end{eqnarray}
in which $X_{\rm e}\equiv {2\pi^2 T_{\rm e}}/{\omega_c}$,
\begin{eqnarray}
&&S(z)=J_0(z)-J_2(z),\\
&&G(z)=J_0(z)-J_2(z)-\frac{z}{2}\,\big[3J_1(z)-J_3(z)\big],\label{G-fun}
\end{eqnarray}
$J_k(z)$ being  the  Bessel function of order $k$.
With the help of power expansions and asymptotic expressions of Bessel functions,
we have 
\begin{eqnarray}
&&S(z)\simeq 1-\frac{3}{8}\,z^2,\\
&&G(z)\simeq 1-\frac{9}{8}\,z^2 \label{small-G}
\end{eqnarray}
for $z\ll 1$; and 
\begin{eqnarray}
&&S(z)\simeq \sqrt{\frac{8}{\pi z}}\cos \Big(z-\frac{\pi}{4}\Big),\\ 
&&G(z)\simeq \sqrt{\frac{8}{\pi z}}\Big[ \cos \Big(z-\frac{\pi}{4}\Big)
-z\,\sin \Big(z-\frac{\pi}{4}\Big)\Big]\hspace{0.5cm}\label{large-G}
\end{eqnarray}
for $z\gg 1$.

In the linear limit ($\epsilon_j\rightarrow 0$ and $T_{\rm e}\rightarrow T$) Eq.\,(\ref{nrxxi}) returns to Eq.\,(\ref{sdhi}).
For nonlinear transport Eq.\,(\ref{nrxxi}) shows that 
the SdHO term, which diminishes when temperature $T_{\rm e}\gg \omega_c/2\pi^2$, 
is modulated with an oscillatory factor 
$S(2\pi\epsilon_j)$ by the finite current density through the dimensionless parameter $\epsilon_j$. 
Furthermore, the non-SdHO part of the differential resistivity, which survives the temperature rising as long as  
$T_{\rm e} \ll \varepsilon_{F}$, exhibits strong oscillation with changing $\epsilon_j$, 
having an oscillation amplitude $\Delta r\equiv r_{xx}^{\rm i}-R_{\rm i0}$ as\,\cite{note}
\begin{equation}
\frac{\Delta r}{R_{\rm i0}}= 2\delta^2\, G(2\pi \epsilon_j).
\label{ampl}
\end{equation} 
Since the Dingle-factor coefficient $2\delta^2$ varies smoothly with changing $\epsilon_j$, 
it is apparent that 
the oscillatory period and the maxima and minima positions of differential resistivity
$r_{xx}^{\rm i}$ are determined mainly by the $G(2\pi\epsilon_j)$ function.
For $\epsilon_j > 0.5$, the $r_{xx}^{\rm i}$ oscillates with $\epsilon_j$ 
approximately in a period $\Delta \epsilon_j \approx 1$, having
maxima around $\epsilon_j \approx n-1/8$ ($n=1,2,3,...$)
and minima around $\epsilon_j \approx n+3/8$ ($n=1,2,3,...$).
Nevertheless, the accurate positions of the maxima and minima vary somewhat with $\epsilon_j$
because of the possible change of $2\delta^2$ with $\epsilon_j$ and 
the $\epsilon_j^{1/2}$-dependent coefficient at large $\epsilon_j$.

\begin{figure}
\includegraphics [width=0.43\textwidth,clip=on] {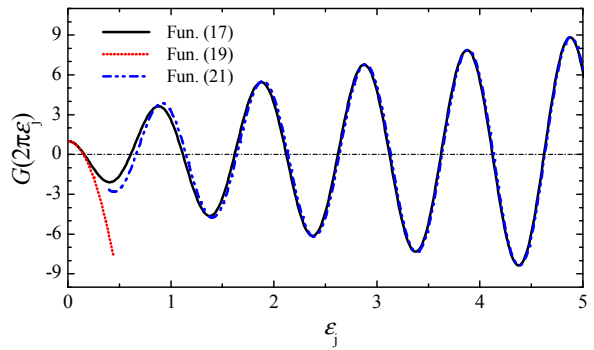}
\caption{(Color online) $G(2\pi \epsilon_j)$ function (\ref{G-fun}) and its 
approximate expression (\ref{small-G}) and (\ref{large-G}) for small and large argument
$z=2\pi\epsilon_j$. 
}
\label{fig1}
\end{figure}
\begin{figure}
\includegraphics [width=0.45\textwidth,clip=on] {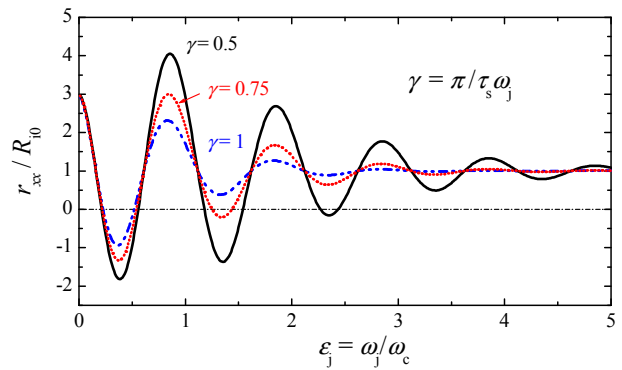}
\caption{(Color online) 
The reduced differential magnetoresistivity $r_{xx}/R_{\rm i0}$ calculated from Eq.\,(\ref{nrxxi})
(excluding the SdHO part) versus $\epsilon_j=\omega_j/\omega_c$
 for several fixed values of $\gamma\equiv \pi/ \tau_s\omega_j  =0.5,\, 0.75$ and 1.
}
\label{fig2}
\end{figure}

The $G(2\pi \epsilon_j)$ function (\ref{G-fun}) is shown in Fig.\,1, together with its approximate
expressions (\ref{small-G}) and (\ref{large-G}) for small and large arguments $z=2\pi\epsilon_j$.
The reduced differential magnetoresistivity $r_{xx}/R_{\rm i0}$ calculated from Eq.\,(\ref{nrxxi})
(excluding the SdHO part), is plotted in Fig.\,2 as a function of $\epsilon_j$ for several fixed
values of $\gamma\equiv (\pi/ \tau_s \omega_j)=0.5,\, 0.75$ and 1.  
This differential resistivity derived here for overlapping LLs, turns out to be in good agreement 
(except for the range of $\epsilon_j<0.5$)
with more general numerical results (covering both separated and overlapping LL regimes) 
of the $B$-sweeping figure in Ref.\,\onlinecite{Lei07-1}
and with the experimental measurement in Ref.\,\onlinecite{WZhang07-1}. 

So far numerical results of theoretical models\cite{Lei07-1,Vav07}
indicate that the maxima and minima of differential resistivity are at positions somewhat lower 
than $\epsilon_j = n$ and $\epsilon_j = n+1/2$ ($n=1,2,3,...$) respectively.
The present analytical expression (\ref{ampl}) with the $G(2\pi \epsilon_j)$ function (\ref{G-fun}) 
[or its asymptotic expression (\ref{large-G})] 
derived from the balance-equation magnetotransport model, predicting a reliable amplitude and phase of the 
oscillatory differential resistance for $\pi\epsilon_j \gg 1$, 
is expected to be useful for an accurate
extraction of temperature-dependent information of the single particle lifetime $\tau_s$ or
the Dingle factor from the current-induced magnetoresistance oscillations in high-mobility
2D electron systems at low temperatures.\cite{Hatke09-4}


\begin{thebibliography}{99}
\bibitem{note} If the oscillation amplitude is defined as $\Delta r\equiv r_{xx}^{\rm i}-r_{xx}^{\rm i}(0)$, then
${\Delta r}/{R_{\rm i0}}= 2\delta^2\, [G(2\pi \epsilon_j)-1]$.
\end{thebibliography}
\end{document}